%% file: main.tex
\documentclass[9pt]{pnas-new}
\templatetype{pnasmathematics} % Choose template 
\title{Lineage EM Algorithm for Inferring Latent States from Cellular Lineage Trees}

\setboolean{displaywatermark}{false}

\author[a,1]{So Nakashima}
\author[b]{Yuki Sughiyama} 
\author[a,b,c,1]{Tetsuya J. Kobayashi}

\affil[a]{Department of Mathematical Informatics, Graduate School of Information Science and Technology, The University of Tokyo, 7-3-1, Hongo, Bunkyo-ku, 113-8654, Japan}
\affil[b]{Institute of Industrial Science, The University of Tokyo, 4-6-1 Komaba, Meguro-ku 153-8505, Tokyo, Japan}
\affil[c]{PRESTO, Japan Science and Technology Agency (JST), 4-1-8 Honcho Kawaguchi, Saitama 332-0012, Japan}

\leadauthor{So Nakashima} 

\authorcontributions{SN, YS, and TJK designed and performed research; SN and TJK analyzed data; SN, YS, and TJK wrote the paper.}
\authordeclaration{The authors declare no conflict of interest.}
\correspondingauthor{\textsuperscript{1} To whom correspondence should be addressed. E-mail: so\_nakashima@mist.i.u-tokyo.ac.jp; tetsuya@mail.crmind.net}

\usepackage[utf8]{inputenc} % allow utf-8 input
\usepackage[T1]{fontenc}    % use 8-bit T1 fonts
\usepackage{hyperref}       % hyperlinks
\usepackage{url}            % simple URL typesetting
\usepackage{booktabs}       % professional-quality tables
\usepackage{amsfonts}       % blackboard math symbols
\usepackage{nicefrac}       % compact symbols for 1/2, etc.
\usepackage{microtype}      % microtypography
\usepackage{amsmath,amssymb}
\usepackage{algorithm,algpseudocode}
\usepackage{booktabs}
\usepackage{comment}
\usepackage{bm}
\usepackage{romannum}

\usepackage[skip=2pt]{caption}

\usepackage{pgfplots}
\pgfplotsset{compat=newest}

\newcommand{\stateSp}{\Omega}

\newcommand{\real}{\mathbb{R}}
\newcommand{\transit}{\mathbb{T}_{\mathrm{F}}}
\newcommand{\gentime}{\pi_{\mathrm{F}}}
\newcommand{\genParameter}{\bm{\theta}^F}
\newcommand{\retroGenParameter}{\bm{\theta}^B}

\newcommand{\empGen}{\pi_{\mathrm{emp}}}
\newcommand{\empGenBW}{\pi_{\mathrm{emp}}^{\mathrm{BW}}}
\newcommand{\empTran}{\mathbb{T}_{\mathrm{emp}}}
\newcommand{\empTranBW}{\mathbb{T}_{\mathrm{emp}}^{\mathrm{BW}}}

\newcommand{\x}{\bm{x}}

\newcommand{\tree}[1]{\mathcal{T}_{#1}}

\newcommand{\retroGen}{\pi_{\mathrm{B}}}
\newcommand{\parameterBW}[1]{\Theta^{(#1)}}

\newcommand{\retroNorm}[1]{Z(#1)}
\newcommand{\retroTran}{\mathbb{T}_\mathrm{B}}
\newcommand{\leftEigen}{u}

\newcommand{\Ecoli}{{\itshape E. coli}}

\newcommand{\cells}{\mathcal{D}}
\newcommand{\eqnref}[1]{Eq. [\ref{#1}]}

\newcommand{\transpose}[1]{{#1}^{\mathbb{T}}}

\newcommand{\secChrono}{S2 and S3}
\newcommand{\secRetro}{S4}
\newcommand{\secProof}{S5}
\newcommand{\secProofRegolous}{S6}
\newcommand{\secBP}{S8 and S10}
\newcommand{\secEstimateLambda}{S7}
\newcommand{\secMstep}{S9}

\newcommand{\secEnv}{S11}
\newcommand{\secDisc}{S12}
\newcommand{\secKalmanImp}{S13}
\newcommand{\secContiValid}{S14}

\newcommand{\secEcoliResult}{S15}
\newcommand{\figEcoliResult}{\ref{fig:lineage_tree_x}}
\newcommand{\relatedWorks}{S1}

\usepackage{color}
\usepackage{soul}
\usepackage{CJKutf8}
\usepackage{ifthen}
\newcommand{\COMM}[2]{{}} %\begin{CJK}{UTF8}{ipxm}
\begin{abstract}
Phenotypic variability in a population of cells can work as the bet-hedging of the cells under an unpredictably changing environment, the typical example of which is the bacterial persistence.
To understand the strategy to control such phenomena, it is indispensable to identify the phenotype of each cell and its inheritance.
Although recent advancements in microfluidic technology offer us useful lineage data, they are insufficient to directly identify the phenotypes of the cells.
An alternative approach is to infer the phenotype from the lineage data by latent-variable estimation.
To this end, however, we must resolve the bias problem in the inference from lineage called survivorship bias.
In this work, we clarify how the survivor bias distorts statistical estimations.
We then propose a latent-variable estimation algorithm without the survivorship bias from lineage trees based on an expectation-maximization (EM) algorithm, which we call \textbf{Lineage EM algorithm (LEM)}.
LEM provides a statistical method to identify the traits of the cells applicable to various kinds of lineage data.
\end{abstract}

\dates{This manuscript was compiled on \today}
\doi{\url{}}

\begin{document}

\maketitle
\thispagestyle{firststyle}
\ifthenelse{\boolean{shortarticle}}{\ifthenelse{\boolean{singlecolumn}}{\abscontentformatted}{\abscontent}}{}

%\input{sections/Intro.tex}
\input{sections/IntroRev.tex}

\input{sections/Lineage.tex}

\input{sections/GrowthBias.tex}

\input{sections/Estimation.tex}

\input{sections/Application.tex}

\input{sections/Discussion.tex}

%\matmethods{Please describe your materials and methods here. This can be more than one paragraph, and may contain subsections and equations as required. Authors should include a statement in the methods section describing how readers will be able to access the data in the paper. 

%\subsection*{Subsection for Method}
%Example text for subsection.
%}

%\showmatmethods{} % Display the Materials and Methods section

\acknow{We greatly thank Yuichi Wakamoto for providing the data of \Ecoli{} for this work and helpful discussion on our results.
This research is supported by JSPS KAKENHI Grant Number JP16K17763, 16H06155, 19H03216, and 19H05799 and by JST Grant Number JPMJPR15E4 and JPMJCR1927 Japan.}

\showacknow{} % Display the acknowledgments section

% Bibliography
\bibliography{references}

\end{document}

%% file: sections/IntroRev.tex
%!TEX root = ../main.tex
\section{Introduction}\COMM{TJK}{Tetsuya have modified introduction}
A population of genetically identical cells is phenotypically heterogeneous and  the heterogeneity is partially inherited over generations~\cite{Kaern2005Stochasticity, RAJ2008Nature, Shahrezaei2008stochastic}.
Such phenotypic variety leads to behavioral individuality of each cell, which, in turn, generates complicated population phenomena.
One example is bacterial persistence in which a fraction of cells in a population survives when the population experiences an antibiotic exposure even though the other fraction dies out~\cite{Bigger1944Treatment,Balaban2004bacterial,Wakamoto2013dynamic, Harmsaaf2016Mechanism, Boxtel2017taking}.
Persistence is also recently recognized relevant to the drug-resistance of cancers \cite{Brock2009genetic, Sharma2010chromatin}.
While the survivors, which are also called persisters, were originally conjectured as dormant and thereby drug-insensitive cells in a population, recent bioimaging analysis revealed that persistence is a more intricate phenomenon, which involves resistant but still growing cells \cite{Balaban2004bacterial,Wakamoto2013dynamic}.
Because drug-resistance is tightly related to the manner how the drug is incorporated into a cell and interferes with the self-replicating process, 
quantitative analysis of persistence requires a characterization of growth states of individual cells and their competition in a population.
%\myHL{(New reff here)}.
More generally, the heterogeneities in the growth speed and the death rate as well as their inheritance from a mother to daughter cells constitute  Darwinian natural selection among the cells.
Such natural selection at the cellular level is also highly relevant to drug-resistances of pathogens and cancers, immunological memories, cell competitions in tissues, and induction of iPS cells  \cite{Brock2009genetic, Sharma2010chromatin, Fisher2017Persistent, Kumar2018Phenotypic, Muller2017Evolutionary}.
Moreover, the population dynamics under selection is integratively determined by the growth states of cells, their statistical property, and their inheritance dynamics over generations. 
Therefore, identification of growth states of cells from data is crucial for predicting and controlling those selection-driven phenomena~\cite{Laessig2017Predicting, Reyes2018Leveraging}.

\begin{figure}[tbh]
  \begin{center}
        \includegraphics[width=0.95 \linewidth]{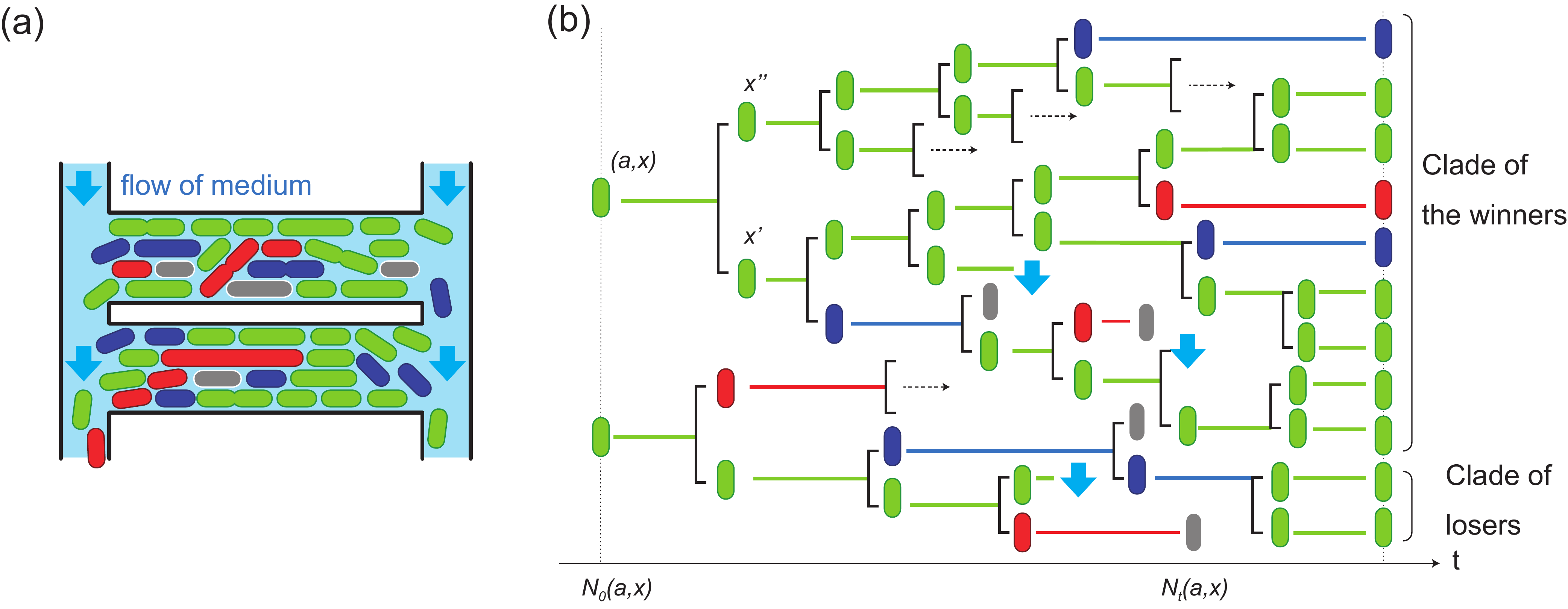}
   \end{center}
    \caption{(a) The outline of the dynamics cytometer~\cite{Hashimoto2016noise}. 
    (b) A schematic representation of a lineage tree obtained by the dynamics cytometer, and an illustration of the survivorship bias.
    %The dynamics of individual cells follows the state-switching and the division time statistics described in (a). 
    The lineage tree is composed of two kinds of information: mother-daughter relationship of the cells and the division times how long each cell took until divides. 
    In this panel, the green state is assumed to divide faster than the blue and red ones. 
    Thereby, the cells with the green state are overrepresented more in the clade of the winner than in that of the losers.
    }
    \label{fig:Intro}
\end{figure}

To determine the growth states and their dynamics, we can take advantage of recent bioimaging and microfluidic technology, which provide abundant but \emph{incomplete} data of the population growth of the cells~\cite{Rowat2009tracking, Wang2010Robust,Hashimoto2016noise}.
%The most widely used device is the mother machine in which a replicating cell is trapped at the bottom of a narrow chamber \cite{Wang2010Robust}.
%By flowing the daughter cells away, we can trace the founder cell at the bottom over tens of generation as long as it is alive, and can obtain samples of the division times over the lineages from multiple founders in parallel \cite{Norman2013memory, taheri-araghi2015cell, Tanouchi2015noisy}. 
Bioimaging enables us to track growing and replicating cells over time and microfluidic devises such as dynamics cytometer substantially boosts our capability of tracking cells beyond hundreds of generations by keeping growing cells in a fixed-sized chamber \cite{Wakamoto2011optimal, Lambert2015Quantifying} (Fig. \ref{fig:Intro} (a)).
The tracked cells in the chamber reconstitute trees of lineages, which contain more detailed information on the mother-daughter relationship of the cells and on the actual competition among the cells in the population (Fig. \ref{fig:Intro} (b)).
%Although such devices offer us useful data, they are still incomplete in the sense that they cannot directly measure all activities in each cell, e.g. the expression levels of all genes. 
%Thus, we cannot \emph{directly} identify the phenotypes of the cells from such data.
From the division patterns in such trees, we may infer cellular growth states as well as their statistical properties and inheritance dynamics.

However, the inference is accompanied by two difficulties.
First, the inference from a tree should be conducted by appropriately handling the branching relationship among the cells in the tree.
This problem has been studied by using the kin-correlation \cite{Hormoz2015inferring, Hormoz2016Inferring}, an algebraic invariance of the lineage tree \cite{Hicks2018Statistical, Hicks2018Maps}, clustering algorithms \cite{Victor2009modified, Failmezger2018clustering}, Monte-Carlo algorithms \cite{Kuzmanovska2017parameter}, and model selection \cite{Kuchen2018long}.
See also Supplementary Information (Section~\relatedWorks{}).
While this problem seems to be addressed by combining these existing estimation techniques in the machine learning, 
this na\"ive anticipations is hampered by the second difficulty.
In the cellular lineage tree, each edge has a different length that reflects the actual division time of the cell.
Therefore, clades of cells replicating faster by chance are represented more than the others in the tree, which inevitably introduces bias in the data sample. 
This is a generalization of the so-called survivorship bias in statistics such that fast-growing clades are overrepresented in a population whereas the slow ones are underrepresented \cite{Wakamoto2011optimal, Hashimoto2016noise, Hoffmann2016nonparametric, Thomas2017making} (Fig. \ref{fig:Intro} (b)).

\begin{figure}[H]
  \begin{center}
        \includegraphics[width=0.7\linewidth]{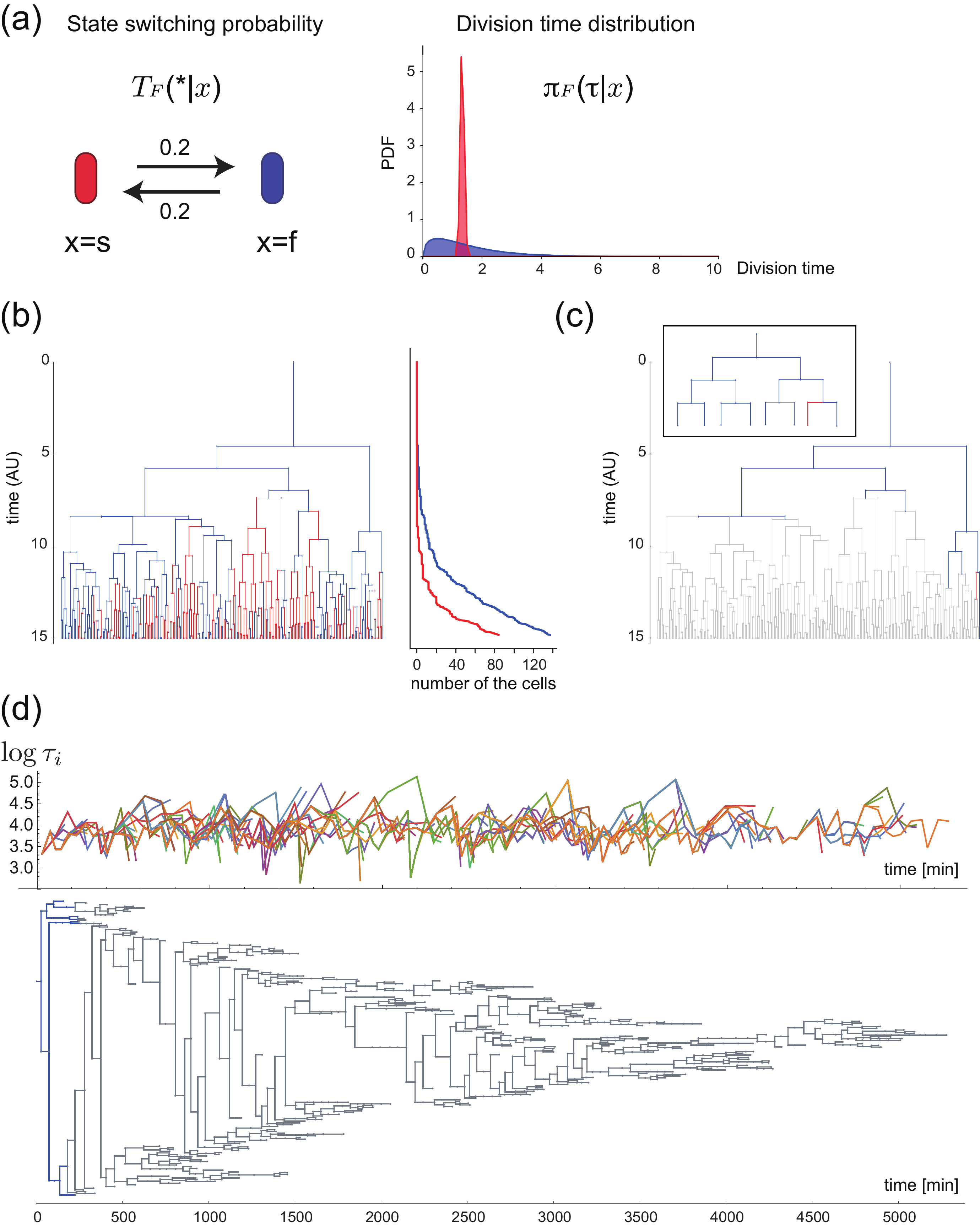}
   \end{center}
    \caption{
    (a) A two-state model to generate a synthetic lineage tree of cells. Red and blue states have a symmetric transition probability $\transit{}(*|\x)$ (left panel). Both states have the same mean division time but the blue state has greater variability in the division time (right panel).
    (b) A synthetic lineage tree obtained by simulating the model in (a). See also Supplementary Information (Section~\secDisc{}). The vertical axis shows time (arbitrary unit). Each edge in the lineage tree represents the time duration between the birth and the division (or death) of the corresponding cells. The horizontal edges show mother-daughter relationships.  
    The right panel shows the cumulative number of the cells in the blue or the red state in the lineage tree over time. The cells in the blue state are much more than those in the red state even though their mean division times are the same.
    (c) A pruned lineage tree of (b) to avoid the survivorship bias, which is shown in color. The cells represented by gray are discarded after pruning. Inset shows the same pruned tree with fixed edge length.
    (d) A lineage tree of \Ecoli{} (F3 rpsL-gfp strain) cells grown with M9 minimum medium supplemented with 0.2\% glucose at $37^\circ$.
    The upper panel in (d) is the time series plot of the division times.
    The lower panel represents the lineage tree of E.~coli.
    The data is adopted from \cite{Hashimoto2016noise} and re-plotted.
    The sub-tree after pruning is also shown in blue. The gray cells are discarded after pruning.     }
    \label{fig:Example}
\end{figure}

To illustrate the overrepresentation, we show a simulated lineage tree of cells with two states. The mean division times of the states are the same but cells in the blue state has greater variability in the division time (Fig. \ref{fig:Example} (a)).
The transition probability between them is symmetric.
In Fig. \ref{fig:Example} (b), each edge in a tree represents the duration between consecutive divisions of a cell and a binary branch is the generation of two daughter cells from a mother cell by its division. 
As shown in Fig. \ref{fig:Example} (b), the cells in the blue state appear more in the tree even though their mean division time is the same as that of the red state.
Thereby, statistical estimators can suffer from sampling bias.
Previous works circumvent this problem by pruning a lineage tree so that each leaf cell has the same number of branching points along the lineage up to the root cell. 
However, as shown in Fig. \ref{fig:Example} (c), such pruning inevitably loses the information about growth statistics of individual cells, which are manifested by the pattern of division times over lineages.

The situation can be even worse when we work on real lineage trees obtained by experiments.
First, the growth states of the cells are not provided as observations but should be inferred from the pattern of divisions of the cells.
Second, as exemplified by the lineage tree of \textit{E.coli} obtained by dynamics cytometer (Fig. \ref{fig:Example} (d)), 
a real tree has frequent terminations of edges because tracking of the corresponding cells is hampered by various reasons. 
In the case of dynamics cytometer, a termination occurs because a cell is carried away from the chamber by the constant flow of medium on the two ends~(Fig. \ref{fig:Intro} (a)).
However, by the virtue of this carry-away, the population size is kept almost constant and we can track a fraction of cells over a hundred generations.
As shown in Fig. \ref{fig:Example} (d), the pruning substantially reduces the number of cells (samples) that we can use for the subsequent inference of cellular states and their division statistics.
If, by contrast, we try to use all the cells in the tree for inference without pruning, we have to appropriately correct the survivorship bias in estimating characteristics of growth states.

Even if we use different culturing devices and cell types, we still face technological limitation upon the number of tracked cells in an exponentially growing population.
Therefore, similar problems should arise when the long-term lineage tracking devices of bacteria are extended to other cells, e.g., cancer cells, immune cells, and iPS cells.

The survivorship bias in an exponentially growing system with states has been analyzed only from recently \cite{Marguet2016uniform, Marguet2017law, Thomas2018Population, Thomas2018analysis}.
Thus, direct inference of cellular growth states without pruning is still immature and we have to establish a method that can estimate and characterize growth states of cells that properly correct the bias.
Furthermore, such method contributes not only to accurate inference but also to estimation of the selection pressure as the strength of the bias \cite{Nozoe2017ingerring, Lambert2015Quantifying}.

%\COMM{SN}{new (sec 追加）}
To address these problems, we first clarify how the survivorship bias distorts statistical estimations depending on the way to collect a sample of cells from the tree (Section~\ref{sec:survivorbias}).
By deriving explicit relations between unbiased and biased estimates, we establish a correction method of the survivorship bias under the condition that the states of the cells are known.
Then, we propose an estimation algorithm of the latent states from lineages based on an expectation-maximization (EM) algorithm, which we call \textbf{Lineage EM algorithm (LEM)} (Section~\ref{sec:LEM}).
Our LEM works under a wide variety of the circumstances without regard to the shape of the lineages nor to the properties of the models in contrast to the previous researches assuming them.
LEM is therefore applicable to the lineage data obtained both by the dynamics cytometer and by the mother machine~\cite{Wang2010Robust} (See also Section~\ref{sec:survivorbias}).
We verify the effectiveness of LEM by using synthetic data (Section~\ref{sec:validation}).
Finally, we apply LEM to a lineage tree of \Ecoli{}, and identify a latent three-dimensional continuous state, one component of which encodes the information on the inheritance of the states and the replicative capabilities over generations.
%See Supplementary Information (Section \secEcoliResult{}).
%The inferred dynamics suggests that the homeostasis and heterogeneity of the division times are controlled with multiple time scales.
 
%However, the division patterns in the partially observed tree still embrace information about how the growth states of the \texitit{E.coli} cells are stochastically determined and inherited over generations.

%In the lineage tree, we can observe continuation, expansion, and extinction of clades of cells, which are induced by heterogeneous growth dynamics and random removals of cells.

%However, fast-growing clades are overrepresented compared with slow-growing ones in the tree, which inevitably bias statistical estimators towards those of fast-growing ones.

%More specifically, the depth of the tree after pruning is constrained by the minimum length of tracked lineages.
%The minimum length cannot be very large not only for this specific example but also for other cells because of technological limitation of tracking all cells in an exponentially expanding population.

%% file: sections/Lineage.tex
%!TEX root = ../main.tex

\section{Statistical modeling of state-switching and division}

%\subsection{A model of lineage with state switching}
\label{sec:Method-Model}

\begin{figure}[tbh]
  \begin{center}
        \includegraphics[width=0.75 \linewidth]{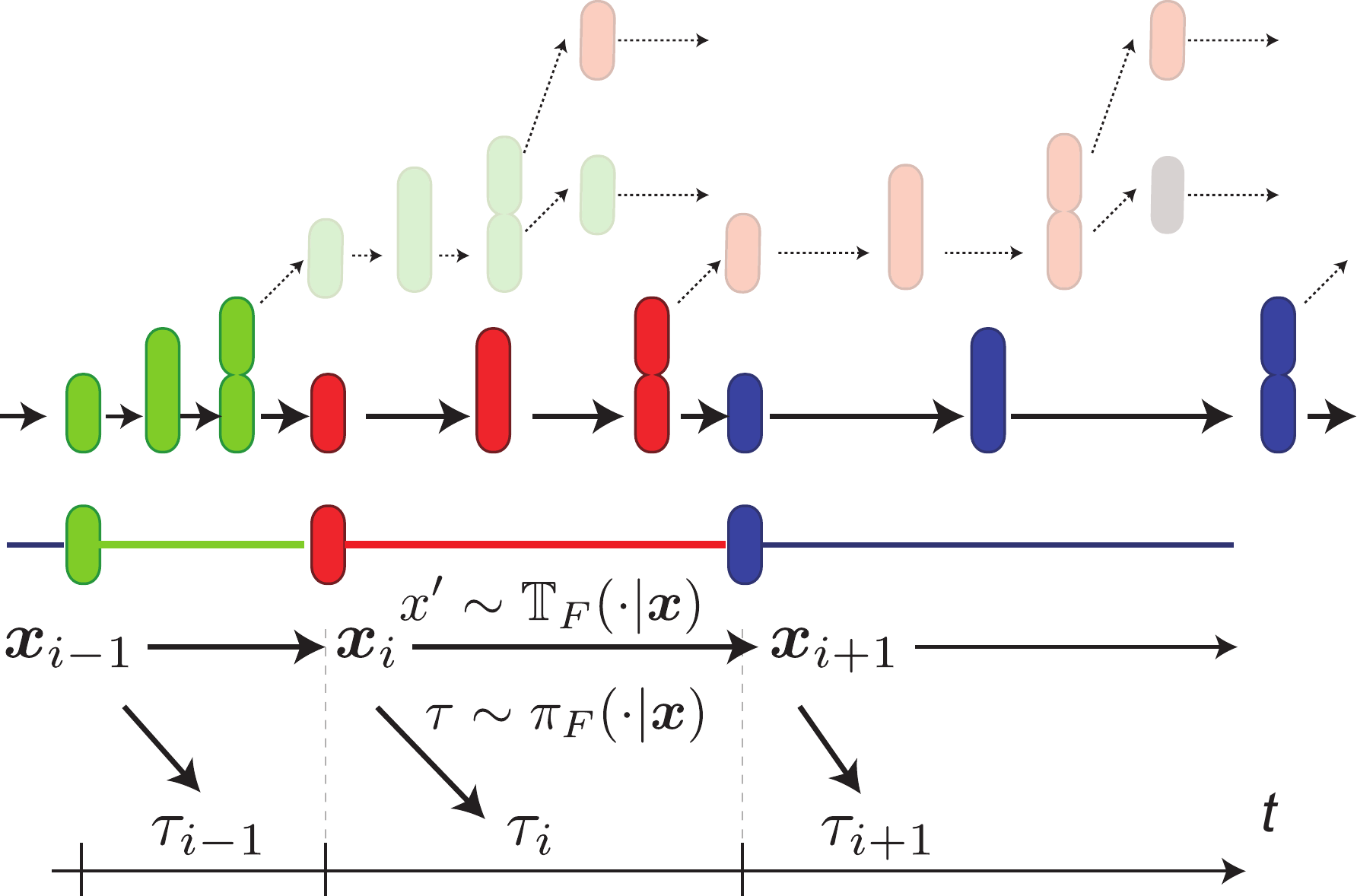}
   \end{center}
    \caption{
    Schematic diagram of stochastic divisions and associated state transitions along a cellular lineage. 
    The colors of cells represent their states. Transparent cells indicate daughter cells that are not in the lineage we are focusing on. 
    All daughter cells are assumed to follow the same statistical laws as those in the focused lineage.
         }
    \label{fig:Switching}
\end{figure}

In this paper, we use a variant of the branching process as a model of a proliferating population with state switching (Fig. \ref{fig:Switching}).
We consider the symmetric division upon which a mother cell always turns into two daughters. 
Each cell is supposed to have its state $\x \in \stateSp$ where $\stateSp$ is either discrete or continuous.
%The states $\stateSp$ are assumed to be discrete for simplicity.
%We will later discuss the continuous state model.
For simplicity, we assume that, upon the division of the mother cell, each daughter cell switches its state stochastically.
However, we note that all the results in this paper is applicable to the cases where the number of the daughters is not necessarily two.
%and those where a lineage obtained by the mother machine. 
The state-switching of a daughter cell is assumed to be dependent on the state of the mother but independent of the state switching of its sister cell.
Then, the probability to change the state from $\x$ to $\x'$ is given by a transition matrix $\transit{}(\x'|\x)$, where $\sum_{\x'} \transit{}(\x' |\x) = 1$ (Fig.~\ref{fig:Switching}).
For notational simplicity, we use $\sum_{\x'\in \stateSp}$ instead of $\int_{\x'\in \stateSp} d{\x'}$ even for $\stateSp$ being a continuous state space. 
The division time $\tau$, the duration time between consecutive divisions, is dependent on the state $\x$ of the cell, the probability distribution of which is denoted by $\gentime{}(\tau|\x)$ (Fig.~\ref{fig:Switching}).
We note that $\transit{}(\x'|\x)$ and $\gentime{}(\tau|\x)$ define a multi-type age-dependent branching process (Fig. \ref{fig:Intro} (b)) \cite{Harris1963theory}, whereas they also constitute a continuous semi-Markov process if one of the daughter cells is ignored (Fig.~\ref{fig:Switching}) \cite{sughiyama2018fitness}.
See also Supplementary Information (Section \secChrono{}).
In general, the state $\x$ of a cell should be characterized as a point or a trajectory in the high dimensional state space consisting of the abundance of intracellular metabolites and molecules in the cell.
However, the state to be inferred in this work can be its low dimensional projection being relevant for the division time,
because any two states, $\x$ and $\x'$, that give the same division statistics as $\gentime{}(\tau|\x)=\gentime{}(\tau|\x')$ cannot be distinguished by the inference only from the data of $\tau$.
Note that, in principle, we can employ other quantities for the inference if provided.

%% file: sections/GrowthBias.tex
%!TEX root = ../main.tex
\section{Correction of the survivorship bias in estimation of state-switching and division dynamics }
\label{sec:survivorbias}

\begin{figure*}[bt]
  \begin{center}
    \includegraphics[width=\linewidth]{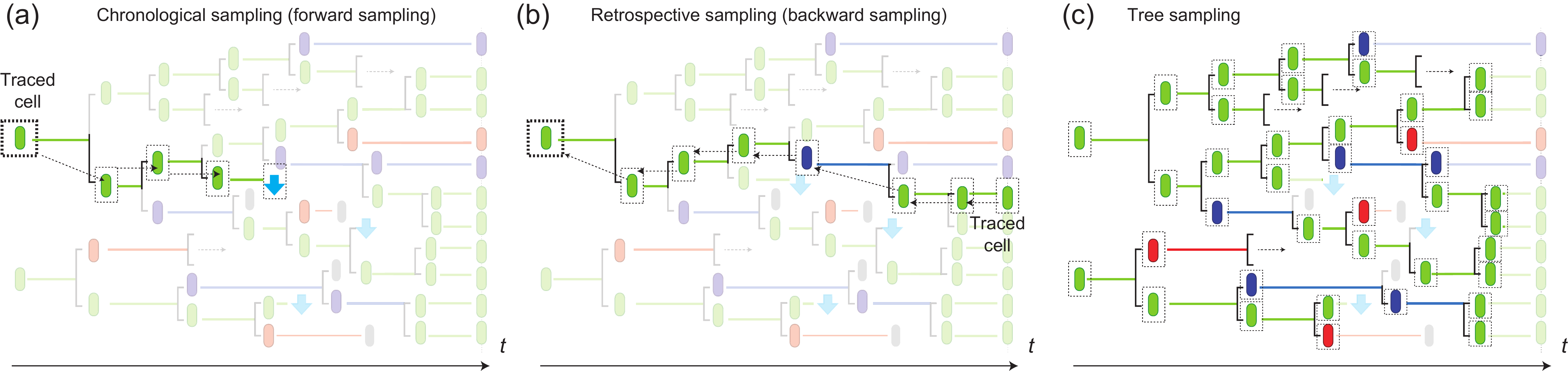}
    \caption{The chronological (a), retrospective (b), and tree (c) samplings are illustrated by using lineage trees. 
    The cells sampled from the tree in each sampling are designated by dashed squares.
    (a) In the chronological sampling, we choose a cell to trace at the beginning of an experiment. Typically, the cell at the bottom of a chamber is chosen in the case of the mother machine. 
    Then, the cell is traced chronologically until we can no longer trace it by either cell death or other reasons.
       We can effectively obtain the chronological sample of a lineage from a lineage tree by tracing the root cell and by randomly choosing one of the two daughter cells upon division.
(b) In the retrospective sampling, we choose a cell to trace randomly at the end of an experiment. Then, we trace the cell retrospectively to its ancestor cells. Because we choose the cell from the survived population, the retrospective lineage cannot be terminated either by cell death or by out of the observation frame. The length of the retrospective lineage is therefore the same as that of the experiment. 
(c) In the tree sampling, we sample all the cells but the leaves, the division times of which were not observed, e.g, by the termination of experiment or by the cell death. 
}
    \label{fig:Method}
  \end{center}
\end{figure*}

%\COMM{SN}{new}
%As the first step to propose a latent-variable estimation without survivorship bias, we clarify how the statistical estimations are distorted depending on the way to collect a sample of cells from the tree. 
If the states of cells are known or experimentally observed, the division time statistics and the state-switching probability, $\gentime{}(\tau|\x)$ and $\transit{}(\x'|\x)$, may be empirically estimated by the histogram of $\tau$ of the cells with $\x$ and by counting the number of the state-switching from $\x$ to $\x'$ in a given data set, respectively:
\begin{align}
  \label{eq:defEmp}
  \empGen^{\cells}(\tau|\x) &:= \frac{1}{|\cells_{\x}|} \sum_{i \in \cells_{\x}} \delta(\tau - \tau_i),\\
    \label{eq:defEmpTran}
  \empTran^{\cells}(\x'|\x) &:= \frac{\text{The number of the transitions from $\x$ to $\x'$}}{\text{The number of the transitions from $\x$}},
\end{align}
where the symbol $|A|$ denotes the cardinality of a finite sample point set $A$, and $\tau_i$ is the division time of the cell $i$.
$\cells$ is the set of all cells used for the estimation, i.e., a data sample, and $\cells_{\x} \subset \cells$ is the subset of the cells with the state $\x$.
$\empTran$ and $\empGen$ may converge for a sufficient large number of cells in $\cells$.
However, the converged distributions are dependent on the way how the cells in $\cells$ were sampled (Fig.~\ref{fig:Method}), and can be substantially biased thereby.

%%%%%%%%%%%%%%%%%%%%%%%%%%%%%%%%%%%%%%%%%%%%%%%%%%%%%%%
\subsection{Chronological sampling and forward process}
Tracking a dividing single cell under a constant condition is the most straight forward way to obtain a data sample of the state-switching events and the division times.
The popular measurement system is the mother machine with which we can trace a cell located at the bottom of a chamber \cite{Wang2010Robust}.
Because the cell to be observed is determined at the beginning of an experiment and its lineage is traced chronologically by ignoring one of the sibling cells at each division, the state-switching and the division dynamics obtained in this way is characterized by the semi-Markov stochastic process with $\gentime$ and  $\transit$ (Fig.~\ref{fig:Switching}).
See also Supplementary Information (Section \secChrono{}).
Thereby, $\empGen$ and $\empTran$ converge to $\gentime$ and $\transit$, respectively, for a large sample size.
We specifically call this type of sampling the chronological sampling and the dynamics generated by $\gentime$ and $\transit$ the forward process \cite{Lambert2015Quantifying, Hashimoto2016noise, sughiyama2018fitness}. 
We can also effectively obtain a chronologically sampled lineage from the tree by using the weighting technique proposed in \cite{Nozoe2017ingerring}.

Even with its straight forward interpretation, the chronological sampling has some drawbacks in terms of the estimation.
First, the observation should be terminated by the death of the tracked cell (Fig. \ref{fig:Method} (a)), which limits the size of the data sample and the length of the lineages especially when the cells are cultured in a harsh condition.
Second, the tracked cells may be exposed to a disturbed environment, because the bottom of the chamber is far from the flowing fresh medium. 
Finally, the chronological sampling does not directly observe the selection process induced by the different replication speeds of the cells in a population. 

%%%%%%%%%%%%%%%%%%%%%%%%%%%%%%%%%%%%%%%%%%%%%%%%%%
\subsection{Retrospective sampling and retrospective process}
These problems can be resolved by using the retrospective sampling of a cell lineage from a proliferating population observed by the dynamics cytometer (Fig. \ref{fig:Method} (b)) \cite{Hashimoto2016noise}.
In the dynamics cytometer, a population of cells is cultured in a more spacious chamber that can accommodates hundreds of the cells, and a cellular lineage tree can be reconstituted from the observed movie.
By sampling a cell from the survived cells in the tree, we can always obtain a cell lineage with the same length of the experiment so long as the cell population rather than a cell does not extinct \cite{Wakamoto2011optimal, Hashimoto2016noise}. 
However, the cells in this retrospective lineage are subject to the survivorship bias, because the lineage is sampled from survived cells.
Thereby, $\empGen$ and $\empTran$ converge to $\retroGen$ and $\retroTran$, which are different from those of the forward process, $\gentime$ and $\transit$.

In order to correct the survivorship bias, in this work, we have proved that $\retroGen(\tau|\x)$ is exponentially biased from $\gentime(\tau|\x)$ as
\begin{align}
  \label{eq:defRetro}
  \retroGen(\tau|\x) &= \frac{2\gentime(\tau|\x) e^{- \lambda \tau}}{\retroNorm{\x}},
\end{align}
where $\lambda$ is the population growth rate of the cells, and $\retroNorm{\x}$ is a normalization factor \cite{sughiyama2018fitness}. See Supplementary Information for the proof (Section \secRetro{}).
This is an extension of Wakamoto \textit{et al.} 2011 \cite{Wakamoto2011optimal} in which the states of the cells were not considered.
We also have derived that $\retroTran(\x'|\x)$ is  biased from $\transit(\x'|\x)$ as
\begin{align}
 \label{eq:defRetroTran}
\retroTran(\x'|\x) = \frac{\leftEigen(\x')  \transit(\x'|\x) \retroNorm{\x}}{\leftEigen(\x)},
\end{align}
where $\leftEigen$ is the left eigenvector associated with the largest eigenvalue of the matrix
\begin{align}
M(\x'|\x) := \transit(\x'|\x)\retroNorm{\x}.
\end{align}
See \cite{sughiyama2018fitness} for the derivation.
This is also an extension of our previous work \cite{Sughiyama2015pathwise} in which the division time was not considered.
$\retroTran$ and $\retroGen$ together define a semi-Markov process of $\x$, which, by construction, asymptotically generates the retrospective cell lineage.
Thus, we call this process the retrospective process. See Supplementary Information (Section \secRetro{}) for the details about the retrospective process.

\eqnref{eq:defRetro} shows that the correction of the bias in $ \retroGen$ requires the population growth rate $\lambda$, which is easily estimated in the dynamics cytometer experiment.
From lineage trees obtained by the dynamics cytometer experiment, we can estimate $\lambda$ as the ratio of the number of the flown cells from the chamber per unit time to that of the whole cells in the chamber.
See also Supplementary Information (Section~\secEstimateLambda{}).
On the other hand,  \eqnref{eq:defRetroTran} indicates that the correction of $\retroTran$ necessitates $\leftEigen$, which can be neither directly observed nor easily estimated. 
This fact limits the use of the retrospective sampling for estimating the cellular state $\x$ and its dynamics.
In addition to this limitation, another problem shared by both chronological and retrospective samplings is that only a lineage of the tracked cell is used for the estimation, which requires quite a long-term tracking to obtain a sufficiently large number of sample points, i.e., the cell divisions and the state-switching events.
In the case of the dynamics cytometer, especially, it seems a huge waste of the data points to abandon the information of the cells being in the tree but out of the tracked lineage.

%%%%%%%%%%%%%%%%%%%%%%%%%%%%%%%%%%%%%%%%%%%%%%%%
\subsection{Tree sampling: estimation from the whole cells in the lineage tree}
These problems can be resolved by the tree sampling in which we use all the cells but the leaves in the lineage tree for estimation (Fig. \ref{fig:Method} (c)).
Here, the leaves correspond to the cells in the tree, the division times of which were not observed, e.g., by the termination of the experiment or flown out from the chamber. 
Yet to be clarified is the bias in the estimation introduced by using the sample obtained in this way.
By employing the many-to-one formulae of the branching process~\cite{Hoffmann2016nonparametric, Marguet2016uniform}, we have proven in this work that $\empGen$ converges to $\retroGen$, whereas $\empTran$ does to $\transit$.
See Supplementary information for the proof (Section \secProof{} and \secProofRegolous{}).
The converged distributions of the chronological, retrospective, and tree sampling are summarized in Tab. \ref{table:threeProcesses}.
%\COMM{SN}{new}
%We also checked this convergence property numerically.
%See Supplementary information (Section \secDisc{}).

Owing to the direct convergence of $\empTran$ to $\transit$ in this tree sampling, we can circumvent the difficulty of reconstructing $\transit$ from $\retroTran$, while enjoying the large number of the sample points in the tree.
Thus, the tree sampling is more efficient than the other samplings.
%\COMM{SN}{new}
The correction of the bias requires the population growth rate $\lambda$.
We can obtain $\lambda$ in two ways. 
The first way is to estimate $\lambda$ from the dynamics cytometer experiment as we mentioned before.
The second way is to calculate $\lambda$ analytically from the estimated $\transit$ and $\retroGen$.
See also Supplementary Information (Section~\secEstimateLambda{}).

\begin{table}[htb]
\centering
\caption{Comparison of the converged distributions obtained by the chronological, the retrospective, and the tree samplings.}
\label{table:threeProcesses}
\begin{tabular}{|l|c|c|c|}
\hline
 & chronological  & retrospective & tree \\ \hline
Division time & $\gentime$ & $\retroGen$ & $\retroGen$ \\ \hline
State switching & $\transit$ & $\retroTran$ & $\transit$ \\ \hline
\end{tabular}
\end{table}

%% file: sections/Estimation.tex
%!TEX root = ../main.tex
\section{Estimation of latent states from a lineage tree}
\label{sec:LEM}

In the preceding section, we have clarified the converged distributions for different samplings under the assumption that the states of the cells as well as the division times are experimentally observed.
However, the information of the states of the cells may not always be accessible.
Even when we observe the expression of several genes over lineages, such genes may not be sufficiently relevant to the determination of the division times, because the division time is generally a consequence of the complicated interactions of intracellular genetic and metabolic networks.
Moreover, even if we could observe the high dimensional state over a lineage, 
we would have to make it interpretable by finding the low dimensional relevant representation of the states to the division times; which generally requires a huge computational cost. 

Such problems can be handled by inferring the effective states of the cells based only on the division time observations.
By extending the EM algorithm for the hidden Markov models \cite{Christopher2016pattern} to a branching tree with hidden states, we construct an algorithm, \textbf{Lineage EM algorithm (LEM)}, for estimating the latent states of the cells in a lineage tree and their dynamic laws. 
To this end, we introduce the following parametric models with  discrete or continuous state-spaces, which enable us to employ well-established statistical methods, e.g. maximum likelihood estimation (MLE), for the estimation.

%%%%%%%%%%%%%%%%%%%%%%%%%%%%%%%%%%%%%%%%%%
\subsection{A parametric discrete state-space model}
For a discrete state-space model, we assume that $\gentime$ belongs to an exponential family \cite{Christopher2016pattern}.
The exponential family includes a broad range of probability distributions such as the gamma-distribution and the log-normal distribution, which have been commonly used for fitting the division time distributions of microbes \cite{Rubinow1968maturity, Wakamoto2011optimal}. 
By assuming a parametric model, the estimation of $\gentime(\tau|\x)$ is reduced to that of the parameter set of the model.
The gamma distribution is a common choice of the parametric model of the division time distribution:
\begin{align}
  \mathbb{P}_{G}(\tau; \bm{\theta}) = \frac{b^{a}}{\Gamma(a)} \tau^{a - 1} e^{- b \tau},
\end{align}
where $\Gamma(a)$ is the gamma function and $\bm{\theta} := (a, b)$, $a$ and $b$ of which are the shape and rate parameters, respectively. 

Then, the division time distribution for the forward process $\gentime(\tau|\x)$ is represented by a $\x$-dependent parameter set $\genParameter_{\x} = (a_{\x}, b_{\x})$ as
\begin{align}
  \gentime(\tau|\x)= \mathbb{P}_{G}(\tau | \genParameter_{\x}).
\end{align}
When $\gentime(\tau|\x)$ is a gamma distribution, so is  $\retroGen$ with a different parameter set, $\retroGenParameter_{\x}$, as
\begin{align}
  \retroGen(\tau|\x)= \mathbb{P}_{G}(\tau| \retroGenParameter_{\x}).
\end{align}
Thereby, we can covert $\retroGenParameter_{\x}$ to $\genParameter_{\x}$ via Eq. [\ref{eq:defRetro}] after estimating $\retroGenParameter_{\x}$. 
On the other hand, the state-switching can be straight-forwardly represented by the components of the matrix, $\transit$.

%%%%%%%%%%%%%%%%%%%%%%%%%%%%%%%%%%%%%%%
\subsection{A parametric continuous state-space model}
Suppose that the continuous state space $\Omega$ is $k$-dimensional Euclidean as $\Omega \subseteq \real^k$.
Because the estimation by considering all the possible dynamics in a continuous state-space is unfeasible, we here adopt a linear diffusion dynamics for the state-switching, $\transit$, which is characterized by a $k \times k$ matrix $\bm{A}$ as
\begin{align}
  \label{eq:LDSstate}
  \x' = \bm{A}\, \x + \bm{w},
\end{align}
where $\x$ and $\x'$ are the states of a mother and its daughter cell, respectively. 
$\bm{w}$ is a multidimensional Gaussian random variable with a mean vector $\bm{0}$ and a diagonal covariance matrix $\bm{\Sigma}_{\bm{w}}$.

The retrospective distribution of the division time, $\retroGen$, is also assumed to follow a log-normal distribution: 
\begin{align}
  \label{eq:LDSgentime}
  \log \tau = \bm{C}\, \x + \bm{v},
\end{align}
where $\bm{C}$ is a $1 \times k$ matrix and $\bm{v}$ is a Gaussian random variable with mean $0$ and variance $\bm{\Sigma}_{\bm{v}}$.
In this model, the estimation problem is reduced to estimating parameters $\bm{A}$, $\bm{C}$, $\bm{\Sigma}_{\bm{w}}$, and $\bm{\Sigma}_{\bm{v}}$, simultaneously.
We remark that the expectation of $\log \tau$ is non-zero due to the non-zero expectation of the initial state $\x_{\mathrm{root}}$ at the root of the lineage.
This setting can be interpreted as a linear approximation of a general continuous-state model of $\x$.

\subsection{Lineage EM algorithm}
To obtain LEM, we extend the Baum-Welch algorithm (BW algorithm) to the estimation of $\retroGenParameter_{\x}$ and $\transit$ from a lineage tree.
LEM algorithm iterates two steps, the E-step and the M-step, and updates the parameters until convergence.
Let $\parameterBW{n}$ denote the estimate of the parameters $(\transit, \{\retroGenParameter_{\x} \})$ after the $n$th iteration.
In the E-step, we compute the posterior probabilities of the states for all the pairs of the mother and daughter cells, $\xi_{i,j}(\x, \x')$, conditioned on the currently estimated parameters $\parameterBW{n}$ and observation.
$\x$ and $\x'$ in $\xi_{i,j}(\x, \x')$ are the states of the cell $i$ and its one of the daughter labeled as the cell $j$, respectively.
$\gamma_{i}(\x)$ is the posterior probability of the state of the cell $i$, which is obtained by marginalization as $\gamma_{i}(\x) = \sum_{\x'} \xi_{i,j}(\x, \x')$. 
$\xi_{i,j}(\x, \x')$ and $\gamma_{i}(\x)$ are computed via the belief propagation \cite{Christopher2016pattern}.
The belief propagation recursively computes the posterior distributions efficiently for a graphical model without loops.
LEM belongs to this class, because a tree is loopless.
See Supplementary Information for the detail (Section \secBP{}).
For the continuous state-space model, we can employ the well-established estimation technique of the Kalman filter \cite{Christopher2016pattern}.
In the M-step, the parameters $\parameterBW{n} = (\transit, \{\retroGenParameter_{\x} \})$ is updated so that $\retroGen(\cdot|\x)$ and $\transit(\x' | \x)$ are fitted to the following modification of the empirical distributions, respectively (\ref{eq:defEmp}):
\begin{align}
  \label{eq:defEmpBW}
  \empGenBW(\tau|\x) &:= \frac{1}{\sum_{i \in \tree{\x}} \gamma_{i}(\x)} \sum_{i \in \tree{\x}} \gamma_{i}(\x) \delta(\tau - \tau_i),\\
  \label{eq:defEmpBW2}
  \empTranBW(\x'|\x) &:= \frac{\sum_{i,j} \xi_{i,j}(\x,\x') }{\sum_{i, j, \x'} \xi_{i,j}(\x,\x')},
\end{align}
where $\tree{\x}$ is the set of all non-leaf cells with state $\x$ in the lineage tree, 
and $(i,j)$ in the second equation runs over all the mother-daughter pairs.
These are empirical distributions weighted by the posterior distributions $\gamma_{i}(\x)$ and $\xi_{i,j}(\x,\x')$.
For the details on the fitting process by MLE, see Supplementary Information (Section \secMstep{}).
It is known that each update always increases the likelihood \cite{Christopher2016pattern}.
In the continuous case, we update $\bm{A}, \bm{C}, \bm{\Sigma}_{\bm{w}}$, and $\bm{\Sigma}_{\bm{v}}$ in the same way, that is, update the parameters so that $\retroGen(\cdot|\x)$ and $\transit(\x' | \x)$ are fitted to $\empGenBW(\cdot|\x)$ and $\empTranBW(\x'|\x)$, respectively.
\COMM{SN}{new}
We remark that the computational time of LEM grows linearly with respect to the number of the cells in the lineage \cite{Christopher2016pattern}.

%% file: sections/Application.tex
%!TEX root = ../main.tex
\section{Validation of LEM}
\label{sec:validation}
We check the validity of our LEM by the estimation from synthetic data.
We also give a validation by the estimation from real lineage trees of \Ecoli{} and its biological consequences in the Supplementary Information \secEcoliResult{}.

%%%%%%%%%%%%%%%%%%%%%%%%%%%%%%%%%
\begin{figure*}[bt]
\begin{center}
	\includegraphics[width=\linewidth]{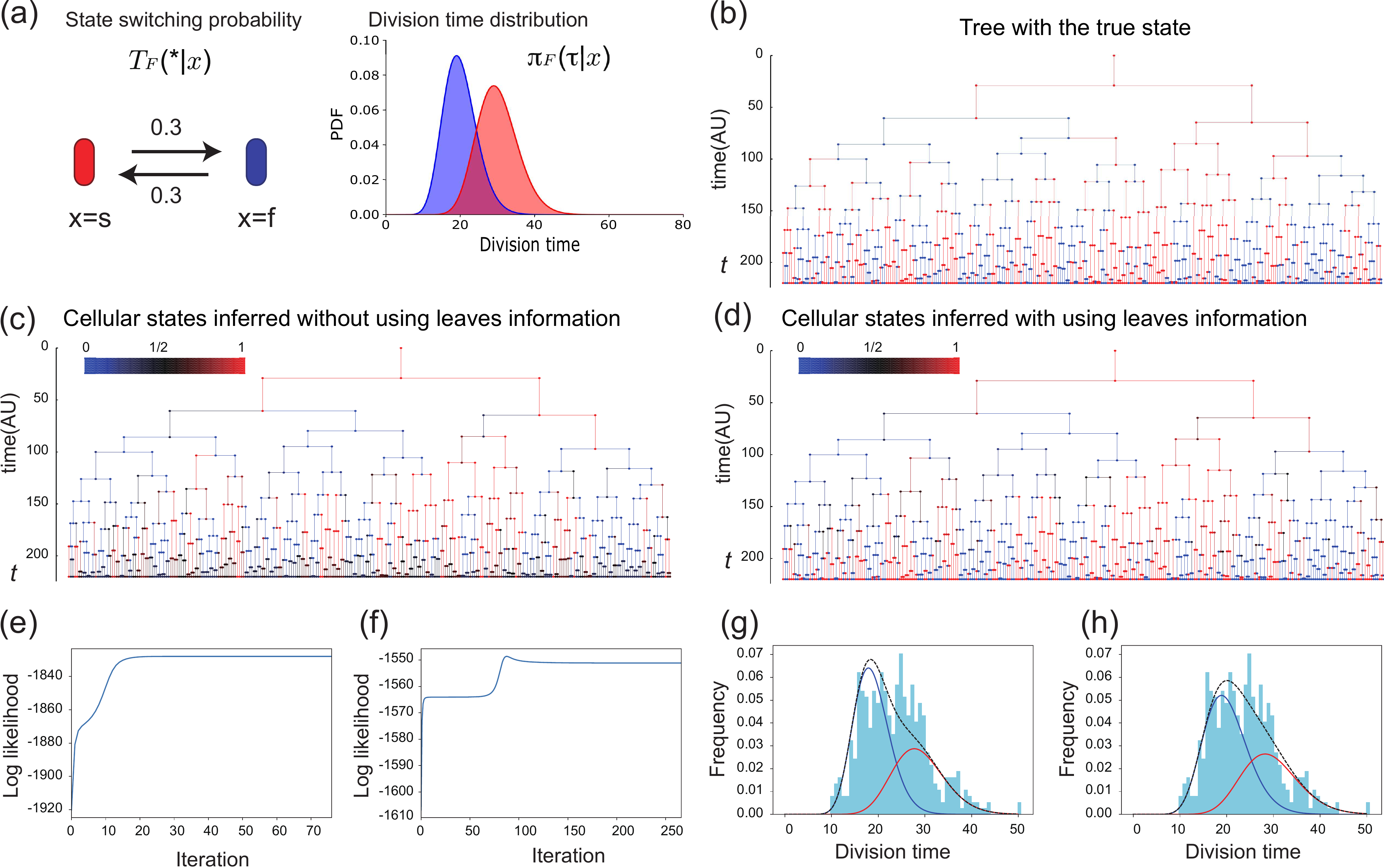}
\end{center}
\caption{A performance evaluation of LEM by comparing a simulated lineage tree of the discrete  model and the corresponding inferred states of the cells with and without using the information of the states of the leaves. 
(a) A schematic diagram of the state-switching dynamics and their division time distributions used for the evaluation. 
Each cell is supposed to have either slowly growing (red) or quickly growing (blue) state, and the transitions between them occur with the probability $0.3$. 
(b) A synthetic lineage tree of the model in (a) obtained by simulating the corresponding branching process.
(c, d) The lineage trees with the latent states inferred from the tree in (b) without (c) and with (d) the information of the actual states of the leaf cells.
The color on a segment indicates the probability that the state of the cell corresponding to the segment is in the red state. 
Red (blue) means that the cell is estimated to be in the red (blue) state with a high probability, whereas 
black means that the estimated state is ambiguous.
(e, f) The convergence of the log-likelihoods when the states of the cells at the leaves are not available (e) and are available (f).  
(g, h) The empirical and the inferred distributions of the division time when we do not know the states of the leaf cells (g) and when we know the states of the leaf cells (h). 
The red and the blue curves show the relative frequencies of the division time of the red and blue states, and the black one is their mixture. The histogram is the empirical distribution of the division times of the cells on the tree.
}
\label{fig:Application1_lineage}
\end{figure*}

\begin{figure*}[bt]
\begin{center}
	\includegraphics[width=\linewidth]{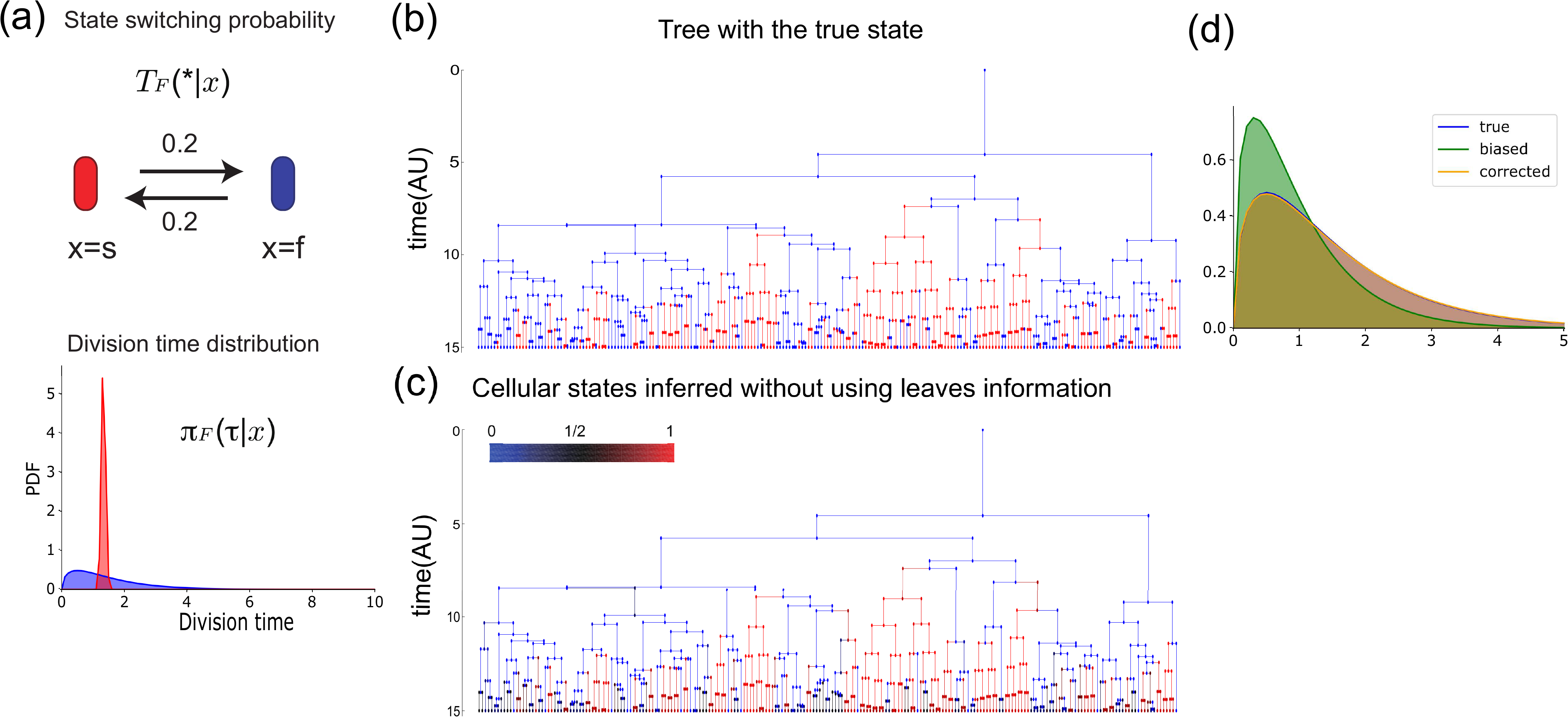}
\end{center}
\caption{A performance evaluation of LEM by comparing a simulated lineage tree of the discrete  model in which the survivorship bias is larger. 
(a) A schematic diagram of the state-switching dynamics and their division time distributions used for the evaluation. This is the same model as Fig.~\ref{fig:Example} (a).
(b) A synthetic lineage tree of the same model as Fig.~\ref{fig:Example} obtained by simulating the corresponding branching process.
(c) The lineage trees with the latent states inferred from the lineage tree without the information of the actual states of the leaf cells.
(d) Correction of the survivorship bias. The blue distribution shows the true distribution of division time (which is almost identical to the corrected  distribution), the green shows the estimated distribution by na\"ive EM-algorithm (without correction of the survivorship bias), and the orange shows the estimated distribution by LEM. %The true distribution and the estimated distribution by LEM conincides.
}
\label{fig:Application1_lineage_latter}
\end{figure*}

%%%%%%%%%%%%%%%%%%%%%%%%%%%%%
\subsection{Validation of LEM with synthetic data sets}
We tested the validity of LEM by numerical experiments of the discrete-state model.
We consider the situation that each cell has two states: a fast-growing ($\x=f$) and slow-growing ($\x=s$) states as depicted in Fig. \ref{fig:Application1_lineage} (a) and obtained a synthetic lineage tree as shown in Fig. \ref{fig:Application1_lineage} (b).
By applying LEM to the lineage tree in Fig. \ref{fig:Application1_lineage} (b), we could recover the states of the cells from the tree as in Fig. \ref{fig:Application1_lineage} (c) without using any state information of the cells.
The states are reliably inferred from the tree containing an experimentally reasonable number of cells, e.g., $500$ cells. See Supplementary Information for the details (Section \secDisc{}{}). 
The states of the leaf cells cannot be inferred in Fig. \ref{fig:Application1_lineage} (c), because the division times of the leaf cells were not observed.
If the state information of the leaves is supplemented for the inference, the accuracy of the estimation is further improved as in Fig. \ref{fig:Application1_lineage} (d).
Such information on the states of the leaves may be obtained by conducting single-cell staining or scFISH \cite{Frieda2016Synthetic}, or scRNA sequencing at the end of the experiment, as assumed in the previous attempts of the state inference from lineage trees \cite{Hormoz2015inferring,Hormoz2016Inferring}. 
The convergence of the log-likelihoods was also checked for both situations (Figs. \ref{fig:Application1_lineage} (e) and (f)). 
%\COMM{SN}{new}
We note that negligible decrease of the likelihood occurs in Fig. \ref{fig:Application1_lineage} (f) due to numerical errors.
We have further compared the empirical and estimated retrospective distributions of the division times (Figs. \ref{fig:Application1_lineage} (g) and (h)) to verify good coincidences between the empirical and the estimated distributions.
Finally, we estimated $\transit$ and $\gentime$ of the model in Fig. \ref{fig:Application1_lineage} (a) for $1000$ times to evaluate the accuracy of our estimation. See Supplementary Information (Section \secDisc{}). 
We observed that the estimation is consistent with the true parameter in total by virtue of the correction of the survivorship bias.
%Similarly, we have also tested LEM for the continuous-model to confirm that LEM also works for that situation. See Supplementary Information (Section \secKalmanImp{}).

\COMM{SN}{new}
We also tested the validity of LEM by using the model shown in Fig.~\ref{fig:Application1_lineage_latter} (a), in which the survivorship bias is larger.
This is the same model as in Fig.~\ref{fig:Example} (a).
By applying LEM to the lineage tree in Fig.~\ref{fig:Application1_lineage_latter} (b), we could recover the states of the cells from the tree as in Fig. \ref{fig:Application1_lineage_latter} (c) without using any state information of the cells.
We estimated $\transit$ and $\gentime$ of the model in Fig. \ref{fig:Application1_lineage_latter} (a)  for $100$ times to evaluate the accuracy of our estimation. See Supplementary Information (Section \secDisc{}).
The survivorship bias is corrected by our method (Fig.~\ref{fig:Application1_lineage_latter} (d)).
In conclusion, our LEM works even for the lineage trees with large survivorship bias.

\COMM{SN}{new}
We note that LEM terminates within a few seconds in this numerical experiments.
See Supplementary information (Section \secEnv{}) for the environment where LEM was executed.

%%%%%%%%%%%%%%%%%%%%%%%%%%%%%%%%%%%%%%%%%%%%%%%%%%
\subsection{Validation of LEM for continous state model}
We checked the validity of LEM for the continuous state model by estimating parameters from synthetic data.
We generated synthetic data as follows.
The lineage tree generated was a perfect binary tree with 10 generations.
The state space is two dimensional as $\stateSp = \real^2$.
The state of the root cell was set to be $\transpose{(1,0)}$.
We adopted the following parameters to generate the synthetic lineage tree:
\begin{align}
\begin{aligned}
\bm{A} &= \begin{pmatrix} 1 & 1 \\ 0 & 1 \end{pmatrix}, & \bm{C} &= \begin{pmatrix} 1 & 1 \end{pmatrix},\\
\bm{\Sigma_w} &= \begin{pmatrix} 0.01 & 0 \\ 0 & 0.01 \end{pmatrix}, & \Sigma_v &= 0.01 .
\label{eq:11.1}
\end{aligned}
\end{align}
We estimated the parameters, i.e., $\bm{A}, \bm{C}, \bm{\Sigma_w}$, and $\Sigma_v$, by LEM.
The initial values of the parameters used in the algorithm were randomly chosen.
See Supplementary Information (Section \secKalmanImp{}). 
We ran the estimation algorithm for 1000 different initial values and adopted the parameters with the largest likelihood.
%\COMM{SN}{new}
LEM terminates in about an hour in this validation.

%Then, by using \eqnref{eq:transparametersLDS}, we obtained transformed parameters $\bm{A}', \bm{C}', \bm{\Sigma_w}', \Sigma_v'$, the coordinate of which coincides with that of the true parameters. 

The estimated parameters became
\begin{align}
\begin{aligned}
\bm{A} &= \begin{pmatrix} 1.040 & 1.093 \\ -0.002 & 0.962 \end{pmatrix}, & \bm{C} &= \begin{pmatrix} 1 & 1 \end{pmatrix},\\
\bm{\Sigma_w} &= \begin{pmatrix} 0.010 & 0 \\ 0 & 0.011 \end{pmatrix}, & \Sigma_v &=  0.008.
\end{aligned}
\end{align}
%Here, we transformed the estimated parameters to compare them with the true parameter correctly.
%See Supplementary Information (Section \secKalmanImp{}).
The estimated parameters are sufficiently close to the true parameters.
This result validates our LEM for the continuous model.

\begin{figure}[htbp]
\begin{center}
\includegraphics[width=0.99\linewidth]{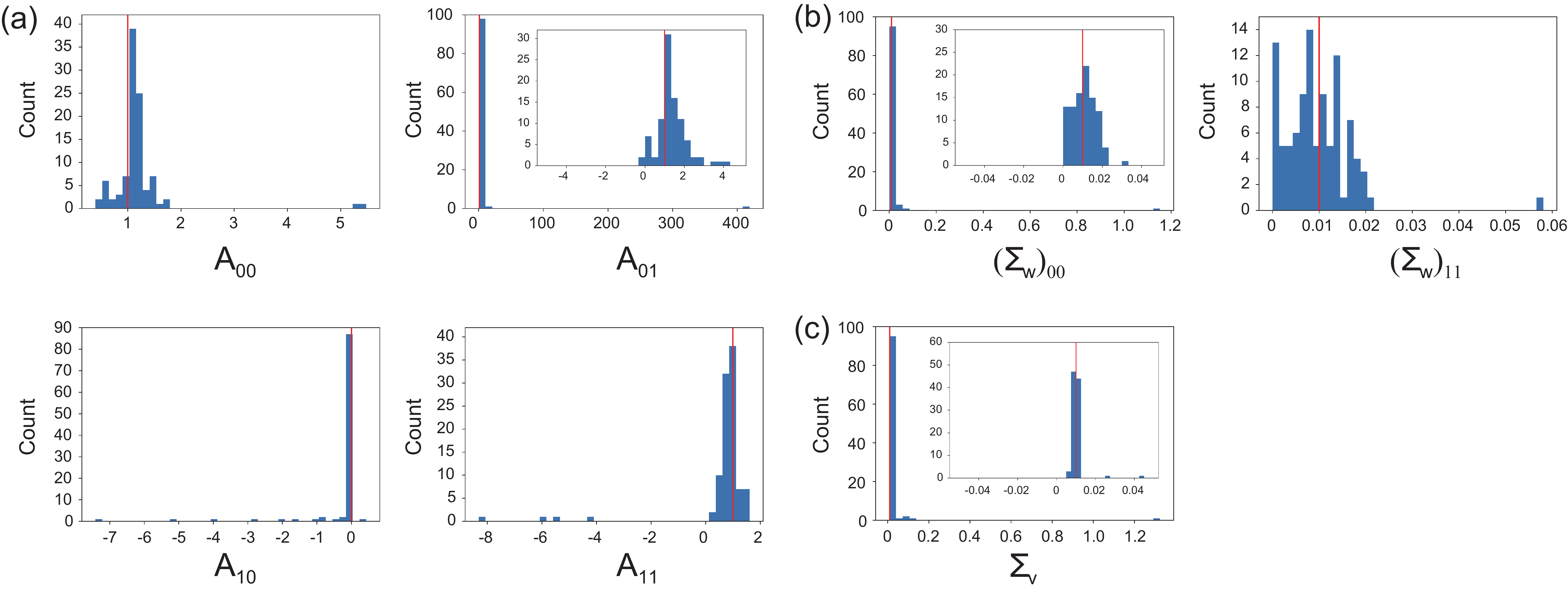}
\caption{The histograms of the estimated values of the parameters, $\bm{A}'$ (a), $\bm{\Sigma_w}'$ (b), and $\Sigma_v'$ (c).
The true value of each parameter is indicated by the red line.
The vertical axes show the counts of the estimated values of the parameters of the $100$ trials.
The insets are the blow-ups of the corresponding histograms around their peaks.
}
\label{fig:checkLDS}
\end{center}
\end{figure}

To see the dependence of the estimated parameters on the sample, we estimated parameters by LEM from 100 synthetic lineage trees generated independently from the same model and the same parameter set (\eqnref{eq:11.1}) as above.
To check the sample-dependency efficiently, we reduced the number of the initial values in the estimation algorithm  from 1000 to 100.
The histogram of the estimated parameter values are summarized in Fig. \ref{fig:checkLDS}.
For each parameter, the estimated value distributed around the true parameters.
Moreover, the estimated $\bm{A}$ conserved the non-zero structure of the true parameter $\bm{A}$.
This result also shows the validity of LEM for the continuous model.
%\COMM{SN}{new}
We also checked that the estimated states by LEM are close to the true states.
See Supplementary Information (Section~\secContiValid{}).

%%%%%%%%%%%%%%%%%%%%%%%%%%%%%%%%%%%%%%%%%%%%%%%%%%

\subsection{Application to E.coli lineage tree}
Finally, in order to demonstrate applicability of our method to real
data, we inferred the latent states of the {\it E. coli} cells in the
lineage trees observed by using the dynamics cytometer in Hashimoto {\it et
al.}~\cite{Hashimoto2016noise} (Fig. \ref{fig:Example} (d)).
By applying LEM of the continuous-state model, in which the dimension
$k$ of the state space $\Omega = \real^k$ was determined by AIC~\cite{Akaike1998information},
we found $k = 3$ to be the dimension of the best continuous model.
LEM terminate in about an hour in this estimation.
%\COMM{SN}{new}
Here, we used the continuous-state model since the discrete-state model cannot explain the correlation between the division time of a mother cell and that of the daughter cells.
See Supplementary information (Section \secEcoliResult{}).
We also obtained the inferred dynamics of the latent state $\x$ over the
lineage tree as in Fig. \figEcoliResult{} and its parameter values as follows:
\begin{align}
\begin{aligned}
\bm{A} &= \begin{pmatrix}-0.731 & 0.438 & 0.032\\ -2.51 & 1.124 & 0.062\\-0.262 & 0.0068 & 1.007 \end{pmatrix}, & \bm{C} &= \begin{pmatrix}1 & 1 & 1\end{pmatrix},\\
\bm{\Sigma}_w&=\begin{pmatrix}0.055 & 0 & 0 \\ 0 & 0.038 & 0 \\ 0 & 0 & 0.016\end{pmatrix},   &\Sigma_v &=0.04.
\end{aligned}\label{eq:infparam}
\end{align}
We verified the statistical significance of the estimated dimension and the
parameter values in
Supplementary Information (Section
\secEcoliResult{}).
Of the three components of the inferred latent state, the first one
has the fastest time-scale of approximately one generation~(Fig.~\figEcoliResult{}~(a)), whereas
the third one changes slowly over generations (Fig. \figEcoliResult{} (c)).
Investigation of the inferred dynamics also indicates that the third
component of a cell encodes a certain amount of information on the
division pattern of its descendants.
See also Supplementary Information (Section
\secEcoliResult{}) for the detail.
This result is consistent with the slow dynamics of certain
intracellular proteins, which might affect the noisy behavior of the
division times and being inherited over generations \cite{Taniguchi2010quantifying}.
Moreover, the inferred states here can be interpreted as the
underlying effective inheritance dynamics that can capture the
correlation of the division times between the mother-daughter pairs.
Such correlation between generations can also be modeled more directly
without the latent state by assuming the conditional dependence of the
division time of a daughter $\tau_0$ on that of the mother $\tau$ as $\gentime(\tau_0|\tau)$ \cite{Kumar2018Phenotypic, Muller2017Evolutionary}.
However, the latent states can offer a way to link the identified states with intracellular physical quantities such as the expressions of candidate proteins.
This link may substantially facilitate our understanding of how the
reproductive capabilities of the cells are determined, regulated, and
inherited as the consequences of the intracellular networks.
%\COMM{SN}{new}

\begin{figure}[t]
\begin{center}
	\includegraphics[width=0.8\linewidth]{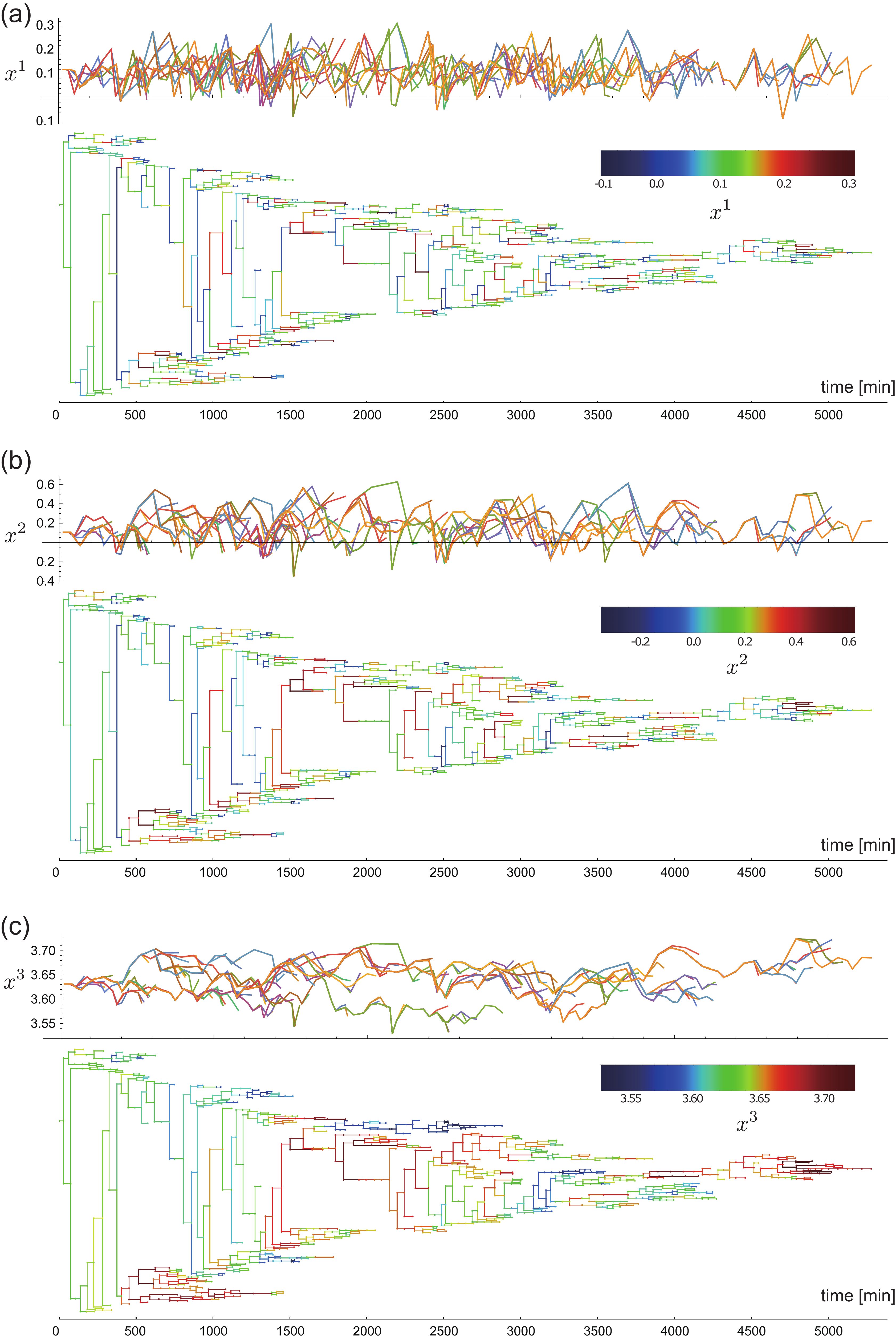}
\end{center}
\caption{
The dynamics of the first (a), the second (b), and the third (c) components of the inferred three-dimensional state depicted as time series (upper panel) and overlaid on the lineage tree (lower panel). 
The color codes over the trees represent the actual values of the components.}
    \label{fig:lineage_tree_x}
\end{figure}

%% file: sections/Discussion.tex
%!TEX root = ../main.tex
\section{Summary and Discussion}
%\subsection{Summary}
In this study, we have derived and proposed LEM, a statistical method to infer the latent states of the cells, the associated state-switching, and division dynamics from lineage tree data.
To this end, we combined the correction method of the survivorship bias with the EM algorithm for trees.
The accuracy and consistency of the method were verified by using the synthetic tree data with two distinct states or by a continuous state model. 
By applying the method to the lineage tree of \Ecoli{}, we have identified the latent low-dimensional states of the cells, which are inherited over a couple of generations at least. 
%\COMM{SN}{new}
%Our LEM is indispensable for distinguishing this low-dimensional state from the effect of the shared environment.
%See also Supplementary Information (Section \secEcoliResult{}).
%The inferred states successfully capture the underlying effective inheritance dynamics of the division times over generations even though the correlation of the observed division times between the mother-daughter pairs is subtle presumably because of the stochastic nature of the cellular replication. 
%

LEM provides a data-driven way to identify and to characterize individual cells in apparently similar yet latently distinct states in a growing population.
Cells in the distinctive modes of the growth, e.g., vegetative and dormant ones, have been identified manually and shown to have different susceptibility to stresses \cite{Balaban2004bacterial, BALABAN2011persistence}.
Recent experimental investigations have further suggested that more subtle differences are still ruling the fates of the cells under the challenge of antibiotics \cite{Wakamoto2013dynamic}.
LEM combined with the dynamics cytometer may play the pivotal roles to investigate the more complicated processes of the cellular natural selections occurring in the populations of bacteria, pathogen, immune cells, and cancer cells \cite{Laessig2017Predicting}.

LEM still leaves room for further improvements that extend its applicability to various problems, some of which may be addressed by using existing techniques of the hidden Markov models.
For instance, we may relax the assumption of the independence of the state-switching between the daughter cells
\cite{Paskin2002simultaneous, Failmezger2018clustering}.
This generalization may be useful when we include the size of a cell as a state, which naturally correlates between the daughters \cite{taheri-araghi2015cell, Susman2018individual}.
We may also extend LEM either to include other experimentally observed quantities than the division times for the estimation of the latent states or to combine the observed quantities as the visible state with the latent states.
The assumption of the linear dynamics in the continuous model or that of the exponential families for the division time distribution can be generalized to incorporate realistic nonlinear dynamics or non-parametric distributions by using Monte-Carlo or ensemble methods at the cost of heavy computational loads \cite{sarkka2013bayesian, Kuchen2018long}.

Moreover, we still have biologically important but theoretically challenging problems: 
One of the problem is the state-dependent death rate of the cells. 
We anticipate that the analysis of the survivorship bias still can be carried over to such situation and conjectures that if a cell dies with a rate $\gamma(\x)$, then the empirical distributions of the generation time converges to 
$\retroGen(\tau \mid \x) = \gentime(\tau \mid \x) e^{-(\lambda + \gamma(\x)) \tau}$.
This $\retroGen$ may be again characterized as the retrospective path from a uniformly chosen cell at the end of an infinitely large lineage. A proof of this conjecture is indispensable for addressing the impact of the antibiotics.
Another is that the feedback from the division time to the latent state transition, which naturally occurs when the latent state is affected how long the next division occurs. 
The feedback inevitably destroys the prerequisite of the BW algorithm that there is no feedback. 
All these problems together with the potential applicability of LEM may open up a new target of the machine learning and the statistics, which will provide quantitative and data-drive ways to address the problems of evolutionary and systems biology.

\section*{Availavility of Materials}
\COMM{SN}{new}
An implementation of LEM is available at \url{https://github.com/so-nakashima/Lineage-EM-algorithm}.